\documentclass[pra,floatfix,onecolumn,nofootinbib,eqsecnum,tightenlines,12pt]{revtex4}
\usepackage{bm}
\usepackage{amsmath}
\usepackage{fancybox}
\usepackage{graphics, graphicx}
\numberwithin{equation}{section}
\renewcommand{\theequation}{\arabic{section}.\arabic{equation}}

\newcommand{\ltwid}{\mathrel{\raise.3ex\hbox{$<$\kern-.75em\lower1ex\hbox{$\sim$}}}}
\newcommand{\gtwid}{\mathrel{\raise.3ex\hbox{$>$\kern-.75em\lower1ex\hbox{$\sim$}}}}

\def\ed{ed.~by~}

\def\chk{{\large{$\bm{\surd}$}}}
\def\cross{{\LARGE{$\bm{\times}$}}}
\def\vy{{\vec y}}

\begin{document}

\title{The quasiclassical realms of this quantum universe\footnote{This is an extended version of a contribution to {\sl Many Worlds?: Everett, Quantum Theory and Reality} edited by  S. Saunders, J. Barrett, A. Kent, and D. Wallace (Oxford University Press, Oxford, 2010) which is the unified proceedings of the conference {\sl Everett at 50}, held at Oxford, July, 2007 and the conference {\sl Many Worlds at 50}, held at the Perimeter Institute, September, 2007.}}

\author{James B.~Hartle}
\email{hartle@physics.ucsb.edu}

\affiliation{Department of Physics,
 University of California,
 Santa Barbara, CA 93106-9530, and \\ Santa Fe Institute, Santa Fe, NM 87501}
\date{\today}

\begin{abstract}
The most striking observable feature of our indeterministic quantum universe is the wide range of time, place, and scale on which the deterministic laws of classical physics hold to an excellent approximation. This essay describes how this domain of classical predictability  of every day experience emerges from a quantum theory of the universe's state and dynamics.

\end{abstract}

\maketitle

\section{Introduction}
\label{sec1}\label{intro}

The most striking observable feature of our indeterministic quantum universe is the wide range of time, place, and scale on which the deterministic laws of classical physics hold to an excellent approximation. What is the origin of this predictable quasiclassical realm in a quantum universe characterized  by indeterminacy and distributed probabilities?  This essay summarizes progress in answering this question both old and new. 

The regularities that characterize the quasiclassical realm are described by the familiar classical equations for particles, bulk matter, and fields, together with  the Einstein equation  governing the regularities of  classical spacetime geometry. 
Our observations of the universe suggest that this  quasiclassical realm extends from a moment after the big bang to the far future and over the whole of the visible volume.  Were we to set out on a journey to arrive in the far future  at a distant galaxy we would count on the regularities of classical physics holding there much as they do here. The quasiclassical realm  is thus a feature of the universe independent of human cognition or decision. It is not a feature that we determine,  but rather one already present that we exploit as information gathering and utilizing systems (IGUSes) acting in the universe. 

So manifest is the quasiclassical realm that it is usually simply assumed in constructing effective physical theories that apply in the late universe. Classical spacetime for instance is the starting assumption for the standard model of the elementary particle interactions. Classical spacetime obeying the Einstein equation is assumed in cosmology to reconstruct the past history of our universe. 

Even formulations of quantum mechanics  assume some part of the universe's quasiclassical realm. 
Copenhagen quantum theory assumed a separate classical physics and quantum physics with a kind of movable boundary between them. Classical physics was the realm of observers, apparatus, measurement outcomes, and spacetime geometry. Quantum physics was the realm of the particles and quantum fields that were being measured.  In the Everett formulations  classical spacetime is  usually assumed in order to define the branching histories which are their characteristic feature.

Classical behavior is not exhibited by every closed quantum mechanical system, only  a small minority of them.  For example, in the simple case of a non-relativistic particle,  an initial wave function in the form of a narrow wave packet may predict a high probability for classical correlations in time between sufficiently coarse-grained determinations of position at a sequence of times. But a generic wave function will not predict high probabilities for such correlations. Classical behavior is manifested only through {\it certain}  sets of alternative {\it coarse-grained} histories and then only for {\it particular} quantum states. 
In particular, we cannot expect the classical spacetime that is the central feature of our quasiclassical realm to emerge from every state in quantum gravity, although it must  from the particular quantum state of our universe.  

This essay summarizes progress in understanding the origin of our quasiclassical realm from a fundamental quantum theory of the universe --- a quantum cosmology.\footnote{This  is not a review of the long history and many different approaches taken to classicality in quantum theory.  Rather it is mostly a brief summary of  the author's work,  much of it with Murray Gell-Mann (especially \cite{GH90a,GH93a,GH07}) within decoherent (or consistent) histories quantum mechanics. The references should be understood in this context. For another approach to classicality in the quantum mechanics of closed systems see \cite{Zur03}. For a different kind of discussion with many references see \cite{Lan05}.}. 
 There are two inputs to this theory:  First, there is the specification of the quantum dynamics (the Hamiltonian in the approximation of classical spacetime.) Second, there is the particular quantum state of our universe. Superstring theory is a candidate for the first input; Hawking's no-boundary wave function of the universe \cite{Haw84a} is a candidate for the second.  An explanation of the quasiclassical realm  from these inputs consists roughly of exhibiting  sets of suitably coarse-grained alternative histories of the universe that have high probabilities for patterns of correlations in time summarized by closed systems of  deterministic classical equations of motion.

The expansion of the universe together with the properties of the strong interactions mean that nuclear densities ($\sim 10^{15} {\rm  g/cm^3}$) are the largest reached by ordinary matter any time past the first second after the big bang. There are nearly $80$ orders of magnitude separating these densities from the Planck density ($10^{93} {\rm g/cm^3}$) characterizing quantum gravity. This large separation in scale permits the division of the explanation of the quasiclassical realm into two parts:  first, the understanding of the origin of classical spacetime in quantum cosmology, and, second, the origin of the classical behavior of matter fields {\it assuming} classical spacetime. 

This division into Planck scale physics and below corresponds to a division in contemporary theoretical uncertainty. But, more importantly, it corresponds to a division in the nature of the explanation of the parts of the quasiclassical realm. As we shall see, the classical behavior of matter follows mostly from the conservation laws implied by the local symmetries of classical spacetime together with a few general features of the effective theory of the elementary particle interactions (e.g. locality) and the initial condition of the universe (e.g. low entropy). By contrast the emergence of classical spacetime involves the specific theory of the universe's quantum state and a further generalization of quantum theory to deal with histories of spacetime geometry. 

These differences should not obscure the point that the explanation of the quasiclassical realm is a unified problem in quantum cosmology.  But  because of them it is convenient to explain the origin of the quasiclassical behavior of matter first and return to the classical behavior of spacetime later. 

This essay is structured as follows:  In Section \ref{ehrenfest} we exhibit a standard text book derivation of classical behavior largely as a foil to the kind of treatment that we aim for. Section \ref{dhqm} sketches the elements of decoherent histories quantum theory.
In Section \ref{oscillators} we consider classicality in a familiar class of oscillator models.
Section \ref{hydro} sketches a general approach to classicality in terms of the approximately conserved hydrodynamic variables. In Section \ref{secondlaw} we briefly discuss the origin of the second law of thermodyamics which is necessary for the understanding of the  origin of the quasiclassical realm as well as being an  important feature of it.  Section \ref{classpacetime} discusses the origin of classical spacetime that is a prerequisite for a quasiclassical realm. Section \ref{whydowe} asks why human observers focus on the quasiclassical realm. Section \ref{copenhagen} considers the Copenhagen approximation to decoherent histories quantum theory  that is appropriate for measurement situations.    Open questions are mentioned in Section \ref{open}. In Section \ref{conclusions} we return to the theme of the connnection between fundamental physics and the quasiclassical realm.  For readers not familiar with it, an appendix gives a bare bones introduction to decoherent histories quantum theory in the notation that we will use.

\section{Classicality from the Ehrenfest Theorem}
\label{ehrenfest}

Standard derivations of classical behavior from the laws of quantum
mechanics are available in many quantum mechanics texts.  One popular
approach is based on Ehrenfest's theorem relating the acceleration of
the expected value of a particle's position to the expected value of the force:
\begin{equation}
m\ \frac{d^2\langle x\rangle}{dt^2} = - \left\langle\frac{\partial
V}{\partial x}\right\rangle\ 
\label{oneone}
\end{equation}
(written here for one-dimensional motion).  Ehrenfest's
theorem is true in general, but for certain states ---typically narrow
wave packets --- we may approximately replace the expected value of the
force with the force evaluated at the expected position, 
thereby obtaining a classical equation of motion for that expected value:
\begin{equation}
m\ \frac{d^2\langle x\rangle}{dt^2} = - \frac{\partial V(\langle x
\rangle)}{\partial x}\ .
\label{onetwo} 
\end{equation}
This equation shows that the center of a narrow wave packet moves on an
orbit obeying Newton's laws.  More precisely, if we make a succession of
position and momentum measurements that are crude enough not to disturb
the approximation that allows \eqref{oneone}  to replace \eqref{onetwo}, then the expected values of
the results will be correlated by Newton's deterministic law.

This kind of elementary derivation is inadequate for the type of
classical behavior that we hope to discuss in quantum cosmology for the
following reasons:

\begin{itemize}
\item {\it Limited to expected values:} The behavior of expected values is not enough
to define classical behavior.  In quantum mechanics, the statement that
the Moon moves on a classical orbit is properly the statement that, among
a set of alternative coarse-grained histories  of its position as a function of time,
the probability is high for those  exhibiting the correlations
in time implied by Newton's law of motion and near zero for all others.
To discuss classical behavior, therefore, we should be dealing with
the probabilities of sets of alternative time histories, not with expected or average
values.

\item {\it Deals only with measurements:} The Ehrenfest theorem
 derivation deals with the results of ``measurements''
on an isolated system with a few degrees of freedom.  However, in
quantum cosmology we are interested in classical behavior over cosmological stretches of space and time, and
over a wide range of subsystems, {\it independently} of whether these
subsystems are receiving attention from observers.  Certainly our observations of the Moon's orbit,  or a bit 
of the 
universe's expansion,
 have little to do with the classical behavior of those systems.
Further, we are interested not just in classical behavior as
exhibited in a few variables and at a few times of our choosing, but  over the bulk of the universe in 
as  refined a description as possible, so that classical behavior
becomes a feature of the universe itself and not a choice of
observers. 

\item {\it Assumes the classical equations follow from the fundamental action:} The Ehrenfest theorem derivation relies on a close
connection between the equations of motion of the fundamental action and
the deterministic laws that govern classical behavior.
But when we speak of the classical behavior of the
Moon, or of the cosmological expansion, or even of water in a pipe, we
are dealing with systems with many degrees of freedom whose
phenomenological classical equations of motion (e.g. the Navier-Stokes equation)  may be only distantly
related to the underlying fundamental theory, say superstring theory.
We need a derivation which
derives the {\it form} of the equations as well as the probabilities that they
are satisfied.

\item {\it Posits rather than derives the variables exhibiting classical behavior:} The Ehrenfest theorem derivation posits the variables
--- the position $x$ --- in which classical behavior is exhibited.  But,
as mentioned above, classical behavior is most properly defined in terms
of the probabilities  of histories.  In a closed system we
should be able to {\it derive} the variables that enter into the
deterministic laws, especially because, for systems with many degrees of
freedom, these may be only distantly related to the degrees of freedom
entering the fundamental action.

\item {\it Assumes classical spacetime:}  The Ehrenfest derivation assumes classical spacetime if only to define the Schr\"odinger equation that underlies \eqref{oneone}. But we aim  at explaining the universe's quasiclassical realms from a quantum cosmology founded on a unified theory of the fundamental interactions including gravity. Generally spacetime geometry will vary quantum mechanically. Classical behavior must therefore be explained not posited. Indeed, we do not  expect to find classical spacetime geometry at the big bang where its quantum fluctuations may be large. Classical spacetime is  part of a  quasiclassical realm,  not separate from it. 
\end{itemize}

Despite these shortcomings, the elementary Ehrenfest analysis
already exhibits two necessary requirements for classical behavior: Some
coarseness is needed
 in the description of the system as well as some restriction on its
initial condition. Not every initial wave function permits the
replacement of (\ref{oneone}) by (\ref{onetwo}) and therefore 
leads to classical behavior;
only for a certain class of wave functions will this be true.
Even given such a suitable  initial
condition, if we follow the system too closely, say by measuring
position exactly, thereby producing a completely delocalized state, we will
invalidate the approximation that allows \eqref{onetwo} to replace \eqref{oneone}
 and classical behavior will not be
expected.  Some coarseness in the description of histories
 is therefore also needed. %

\section{Decoherent histories quantum mechanics}
\label{dhqm}
The conferences to which this article is a  contribution marked 50 years of Everett's formulation of quantum theory. But they were only a year away from marking 25 years of  decoherent (or consistent) histories quantum theory that can be viewed as extension and to some extent a completion  of Everett's work  (e.g. \cite{Gri02,Omn94,Gel94}). Today, decoherent histories is the a formulation of quantum theory that is logically consistent, consistent with experiment as far as is known, consistent with the rest of modern physics such as special relativity and field theory, general enough for histories, general enough for cosmology, and generalizable for  quantum gravity. It may not be the only formulation with these properties but it is the only such we have at present.  Quasiclassical realms are defined through the probabilities of histories of the universe.  Decoherent histories quantum theory is the framework for computing them. 

The basics of decoherent histories quantum mechanics in a classical background spacetime are reviewed briefly in the appendix\footnote{Alternatively see \cite{Har93a} for a tutorial in the present notation.}. We recap the essential ingredients here:
For simplicity we consider a model cosmology consisting of  a closed system of particles and fields in a very large box. The basic theoretical inputs are a Hamiltonian $H$ specifying quantum dynamics in the box and a density matrix $\rho$ specifying the  box's initial quantum state.  Coarse-grained histories are represented by class operators $C_\alpha$. In an operator formulation these are chains of  Heisenberg picture projections at a series of times formed with the aid of $H$  [cf \eqref{twothree}]. In a path integral formulation they can be bundles of Feynman paths $q^i(t)$ in configuration space. 

Probabilities are properties of exhaustive sets of exclusive histories $\{C_\alpha\}$, $\alpha=1,2,3, \cdots$. Decoherence is a sufficient condition for their probabilities $\{p(\alpha)\}$  to satisfy the usual rules of probability theory. The central relation  defining both decoherence and probability  is
\begin{equation}
D\left(\alpha^\prime, \alpha\right) \equiv Tr\,\bigl(C_{\alpha^\prime} \rho
C^\dagger_\alpha\bigr) \approx \delta_{\alpha' \alpha} p(\alpha)  . 
\label{dfnal} 
\end{equation}
The  first equality defines the {\it decoherence functional}. The second defines  decoherence and the probabilities that are predicted from $H$ and $\rho$. A decoherent set of alternative coarse-grained histories is called a {\it realm} for short\footnote{There will generally be families of realms defined by closely related coarse grainings that exhibit classical behavior.  Realms employing slightly different intervals for defining coarse-grained position are a simple example. Thus it  would be more accurate to refer to the quasiclassical realm{\it s} exhibited by the universe rather than the quasiclassical realm and we shall do so from now on.},

In a path integral formulation,  sets of alternative coarse-grained histories can be defined by partitioning fine-grained configuration space paths $q^i(t)$ into exhaustive sets of exclusive classes $\{c_\alpha\}$. A useful transcription of the decoherence functional  \eqref{dfnal} for such coarse-grained histories  on a time interval $[0,T]$ is  
\begin{equation}
D\left(\alpha^\prime, \alpha\right) = \int_{c_{\alpha^\prime}}
\delta q^\prime \int_{c_\alpha} \delta q\, \delta \bigl(q^\prime_f
- q_f\bigr)\ e^{i(S[q^\prime(\tau)] - S[q(\tau)])/\hbar} \rho \left(q^\prime_0,
q_0\right) .
\label{sohdfnal}
\end{equation}
Here, the integrals are over fine-grained paths $q^i(t)$ lying in the classes $c_{\alpha'}$ and $c_\alpha$,  $S[q(t)]$ is the action corresponding to the Hamiltonian $H$, and $\rho(q'_0,q_0)$ is the configuration space representative of the initial density matrix $\rho$.

\section{Oscillator Models}
\label{oscillators}
The oscillator models pioneered in \cite{FV63,CL83,UZ89} and \cite{GH93a} and developed by many others provide an explicitly computable setting for understanding aspects of classicality. The following assumptions define the simplest model. 

\begin{itemize}

\item We consider a single distinguished oscillator of mass $M$, frequency $\omega_0$,  and coordinate $x$ interacting with a bath of other oscillators with coordinates $Q^A$, $A=1,2,\cdots$. The coordinates $q^i$ in \eqref{sohdfnal} are then $q^i=(x,Q^A)$. 

\item We suppose the action to be  the sum of an action for the
$x$, an action for the $Q$'s, and an interaction that is a linear coupling between them. That is, we assume the action has the form.
\begin{equation}
S[q(\tau)] = S_{\rm free} [x(\tau)] + S_0 [Q(\tau)] + S_{\rm int}[x(\tau), Q(\tau)] .
\label{oscaction}
\end{equation}
More specifically, the associated Hamiltonians are
\begin{equation}
H_{\rm free} = \frac{1}{2}(M{\dot x}^2 + M\omega_0^2 x^2) ,
\label{hfree}
\end{equation}
 a similar form with different masses and frequencies for $H_0$, and  
\begin{equation}
H_{\rm int} = x \sum_A  g_A Q^A   
\label{intham}
\end{equation}
for some coupling constancts $g_A$. 

\item We suppose the initial density matrix $\rho$ factors into a
product of one depending on the $x$'s and another depending on the $Q$'s which are often called the ``bath'' or the ``environment'', viz: 
\begin{equation}
\rho\left(q^\prime_0, q_0\right) = \bar\rho \left(x^\prime_0,
x_0\right)\, \rho_B \left(Q^\prime_0, Q_0\right)\ . 
\label{denmatrix}
\end{equation}
We assume that the bath oscillators are in a thermal state $\rho_B$ characterized by a temperature $T_B$. 

\item We restrict attention to a simple set of alternative coarse-grained histories that follow the coordinate $x$ of the distinguished oscillator while ignoring the coordinates $Q^A$ of the bath. The histories of the distinguished oscillator are specified by giving an exhaustive set of exclusive intervals of $x$  at each of a series of times $t_1, t_2, \cdots t_n$.   A coarse-grained history  $c_\alpha$ is the bundle of paths $x(t)$ passing through a particular sequence of intervals  $\alpha\equiv (\alpha_1, \alpha_2, \cdots \alpha_n)$ at the series of times $t_1, t_2, \cdots t_n$. For simplicity we take all the intervals to be of equal size $\Delta$ and the times to be equally separated by $\Delta t$. 
\end{itemize}

Since the bath oscillators are unconstrained by the coarse graining, the integral  over the $Q$'s in \eqref{sohdfnal}  can be
carried out to give a decoherence functional just for coarse-grained
histories of the $x$'s of the form: 
\begin{equation}
D\left(\alpha^\prime, \alpha\right) = \int_{c_{\alpha^\prime}} \delta x^\prime  
\int_{c_{\alpha}} \delta x\, \delta\bigl(x^\prime_f - x_f\bigr)
$$
$$\times \exp
\biggl\{i\Bigl(S_{\rm free} [x^\prime (\tau)]
 - S_{\rm free} [x(\tau)] + W
\left[x^\prime (\tau), x(\tau)\right]\Bigr)/\hbar\biggr\}\,
\bar\rho\left(x^\prime_0, x_0\right) 
\label{xdchfnal}
\end{equation}
where $W [x^\prime(\tau), x(\tau)]$, called the 
Feynman-Vernon influence phase, summarizes the
results of integrations over the $Q$'s. 

In the especially simple case of a cut-off continuum of bath
oscillators and high bath temperature the imaginary part of the influence phase is given by \cite{CL83}:
\begin{equation}
{\rm Im}W[x'(\tau),x(\tau)]= \frac{2M\gamma kT_B}{\hbar} \int^T_0 dt \left(x^\prime(t) -
x(t)\right)^2 
\label{infph-im}
\end{equation}
where  $\gamma$ is a measure of the
strength of its coupling to the bath related to the $g_A$ in \eqref{intham}. ${\rm Im}W$ becomes substantial when $x'(\tau)$ and $x(\tau)$ are very different and the time difference $\Delta t$ is long enough. Then the off-diagonal elements of $D(\alpha',\alpha)$ are exponentially suppressed meaning that the set of alternative histories approximately decoheres [cf \eqref{dfnal}]. Roughly, the coarse-graining time required is 
\begin{equation}
\Delta t \gtrsim t_{\rm decoh} \equiv \frac{\hbar^2}{2M\gamma k T_B \Delta^2} .
\label{decohtime}
\end{equation}
The time $t_{\rm decoh}$ is called the decoherence time \cite{Zur84}.  This is typically very much shorter than typical dynamical time scales, for instance $1/\gamma$.  

The diagonal elements of the decoherence functional \eqref{xdchfnal} are the probabilities $p(\alpha)$ of the individual histories in the set (c.f \eqref{dfnal}). With a little work these can be expressed in the following form \cite{GH93a}:
\begin{equation}
p(\alpha)= \int_{c_\alpha} \delta x \ (\cdots) \exp\left[-\int dt  \left(\frac{M^2}{4\hbar}\right) \left(\frac{\hbar}{2 M\gamma kT_B}\right) E(x(t))^2 \right] w(x_0,p_0) , 
\label{probE}
\end{equation}
the dots denoting factors irrelevant for the subsequent argument. 
Here $w(x_0,p_0)$ is the Wigner distribution for the density matrix of the distinguished particle $\bar\rho$ [cf. \eqref{denmatrix}]  and  $E$ is 
\begin{equation}
E(x(t))\equiv \ddot x +\omega^2 x +2\gamma \dot x  
\label{defE}
\end{equation}
where  $\omega$ is the frequency of the $x$-oscillator $\omega_0$  renormalized by its
interaction with the bath. Eq \eqref{probE} has been organized to show that the factor in front of the imaginary part of the influence phase \eqref{infph-im} appears inversely in the exponent of this relation. 

$E=0$ is the classical equation of motion for the distinguished oscillator. This includes a  frictional force arising from the interaction of the particle with the bath. When the coefficient in front of $E^2$ in \eqref{probE} is large,  the probabilities for histories $p(\alpha)$ will peak about histories that satisfy the classical equations of motion. Thus classical behavior of the distinguished oscillator is predicted. 
The width of the distribution is a measure of thermal and quantum noise causing deviations from classical predictability. 
 
In this simple case, an analysis of the
requirements for classical behavior is straightforward.  
Eq.  \eqref{infph-im} shows that
high values of $M\gamma kT_B/\hbar$ are needed to achieve decoherence.  Put differently, a  strong coupling between the distinguished oscillator and the bath is required if
interference phases are to be dissipated efficiently into the bath.
However, the larger this coupling is,   the smaller the
coefficient in the exponent of \eqref{probE} is, decreasing the size of the exponential
and {\it increasing} deviations from classical predictability.  This is
reasonable: the stronger the coupling to the bath the more noise
is produced by the interactions that are carrying away the phases.  To
counteract that, and achieve a sharp peaking about the classical
equation of motion, $M^2/4\hbar$ must be large. 
That is, high inertia is needed to resist the noise that arises from the
interactions with the bath.

Thus, much more coarse graining is needed to ensure classical
predictability than naive arguments based on the uncertainty principle
would suggest.  Coarse graining is needed to effect decoherence, and
coarse graining beyond that to achieve the inertia necessary to resist
the noise that the mechanisms of decoherence produce.

This derivation of classicality deals genuinely with histories, and is not restricted to measurements. But there is still a close connection between the classical equations and the fundamental action. The variable $x$ which behaves classically was posited, not derived, and classical spacetime was assumed. The progress in relation to the Ehrenfest derviation is summarized in the table below: \\

\noindent\shadowbox{\parbox{6.5in}{
\centerline{Deficiencies of the Ehrenfest Derivation}
\begin{itemize}
\item[\chk] {\it Limited to expected values, but classicality is defined through histories.}  

\item[\chk] {\it Deals only with measurements on isolated subsystems with a few degrees of freedom.} 

\item[\cross] {\it Assumes the classical equations follow directly from the fundamental action.} 

\item[\cross] {\it Posits rather than derives the variables which exhibit classical behavior.}  

\item[\cross] {\it Assumes classical spacetime.} 

\end{itemize}
}}

\section{Quasiclassical Coarse-Grainings, Local Equilibrium,  and Hydrodynamic Equations}
\label{hydro}

Isolated systems evolve toward equilibrium; that is a consequence of statistics. But conserved or approximately conserved quantities approach equilibrium more slowly than others.
These include conserved quantities like energy and momentum that arise from the local symmetries of classical spacetime together with conserved charges and numbers arising from the effective theory of the particle interactions. A situation of {\it local equilibrium} will generally
be reached before complete equilibrium is established, if it ever is. This local equilibrium is characterized
by the values of conserved quantities averaged over small volumes.  Even for systems of modest size, time
scales for small volumes to relax to local equilibrium can be very, very much shorter than the time scale for reaching
complete equilibrium.  Once local equilibrium is established, the subsequent evolution of
the approximately conserved quantities can be described by closed sets of effective
classical equations of motion such as the Navier-Stokes equation.  The local equilibrium
determines the values of the  phenomenological quantities such as pressure and viscosity that enter into these equations and the constitutive relations among them. 

That in a nutshell is the explanation of the quasiclassical realms of matter given classical spacetime. It both identifies the variables in which the quasiclassical realms are defined and the mechanism by which they obey closed sets of equations of motion. To make this more concrete we will review very briefly the standard derivation (e.g. \cite{Fos75,Zub74}) of these equations of motion in a simple model. We follow \cite{GH07} where more detail can be found. In \cite{Hal98,Hal99} Jonathan Halliwell explains why sets of sufficiently coarse-grained histories of these variables decohere and lead to high probabilities for correlations in time summarized by the same equations of motion. 

Consider a system of conserved particles inside a non-rotating box interacting by local short range potentials. Let the density matrix $\rho$ --- possibly pure --- describe the state of the system.  Divide the box up into equal volumes of size $V$ labeled by a discrete index $\vy$. Let  $T^{\alpha\beta} (\vec x, t)$ be the  stress-energy-momentum 
operator in the Heisenberg picture.  The energy density $\epsilon 
(\vec x, t)$ and momentum density $\pi^i (x, t)$ are $T^{tt}(\vec x, t)$ and $T^{ti} 
(\vec x,t)$ respectively.  Let $\nu (\vec x, t)$ denote the number density of the conserved particles. Then define
\begin{subequations}
\label{sixone}
\begin{eqnarray}
{{\epsilon}_V}(\vec y, t) & \equiv & \frac{1}{V} \int_{\vec y} d^3x\, \epsilon (\vec x, t) ,
\label{sixone a}\\
\vec\pi_V (\vec y, t) & \equiv & \frac{1}{V} \int_{\vec y} d^3x
\vec\pi(\vec x, t) ,
\label{sixone b}\\
\nu_V (\vec y, t) & \equiv & \frac{1}{V} \int_{\vec y} d^3x\, \nu (\vec x, t) ,
\label{sixone c}
\end{eqnarray}
\end{subequations}
where in each case the integral is over the volume labeled by $\vec y$. These are the
quasiclassical variables for our model.  We note that the densities in \eqref{sixone}
are the variables for a classical hydrodynamic description of this system --- for example, the variables of the Navier-Stokes equation. 

Were the system in complete equilibrium the expected values of the quasiclassical variables defined from the density matrix $\rho$  could be accurately computed from the effective density matrix 
\begin{equation}
\tilde\rho_{\rm eq} = Z^{-1} \exp[-\beta (H-\vec U\cdot \vec P -\mu N)]\, .
\label{sixthree}
\end{equation}
Here, $H$, $\vec P$, and $N$ are the operators for total energy, total momentum, and total
conserved number inside the box --- all extensive quantities.  The $c$-number intensive
quantities $\beta$, $\vec U$, and $\mu$ are respectively the inverse temperature (in units
where Boltzmann's constant is 1), the velocity of the box, and the chemical potential.  A
normalizing factor $Z$ ensures $Tr(\tilde\rho_{\rm eq})=1$.  In equilibrium the expected values are, for instance, 
\begin{equation}
\langle\epsilon_V (\vec y, t)\rangle \equiv Tr (\epsilon_V (\vec y, t) \rho) \approx Tr (\epsilon_V (\vec y, t) \tilde\rho_{\rm eq})\, 
\label{sixfour}
\end{equation}
Indeed, this relation and similar ones for $\vec\pi_V(\vec x, t)$  and $\nu_V(\vec x, t)$ {\it define} equilibrium.

Local equilibrium is achieved when the decoherence functional for sets of histories of
quasiclassical variables $(\epsilon, \vec\pi, n)$ is given approximately by the {\it local} version of
the equilibrium density matrix \eqref{sixthree}  
\begin{equation}
\tilde\rho_{\rm leq}=Z^{-1} \exp\Bigl[-\int d^3y \, \beta(\vec y, t)\, \bigl(\epsilon_V(\vec y,
t) - \vec u(\vec y, t)\cdot \vec \pi_V\, (\vec y, t) - \mu\, (\vec y, t)\, \nu_V\, (\vec y,
t)\bigr)\Bigr] \ . 
\label{sixfive}
\end{equation}
(The sum over $\vy$ has been approximated by an integral.)
Expected values are given by \eqref{sixfour} with $\tilde\rho_{\rm eq}$ replaced by $\tilde\rho_{\rm leq}$.
The expected values of quasiclassical quantitites are thus functions of the 
intensive $c$-number quantities $\beta(\vec y, t)$, $\vec u(\vec y, t)$, and $\mu(\vec y,
t)$. These  are the local inverse temperature, velocity, and chemical potential respectively. They now vary with time and place as the system evolves toward complete equilibrium.

A closed set of deterministic equations of motion for the expected values of $\epsilon (\vec x,
t)$, $\vec\pi (x, t)$, and $\nu(\vec x, t)$ follows from assuming that $\tilde\rho_{\rm leq}$
is an effective density matrix for computing them.  To see this, begin with the Heisenberg equations for the conservation of the  stress-energy-momentum operator $T^{\alpha\beta}(\vec x, t)$ and the number
current operator $j^\alpha(\vec x, t)$.
\begin{equation}
\frac{\partial T^{\alpha\beta}}{\partial x^\beta}=0\quad , \quad \frac{\partial
j^\alpha}{\partial x^\alpha}=0\, .
\label{sixsix}
\end{equation}
 Noting
that $\epsilon(\vec x, t)=T^{tt} (\vec x, t)$ and $\pi^i (\vec x, t)=T^{ti} (\vec x, t)$,
eqs \eqref{sixsix} can be written in a 3$+$1 form and their expected values taken.  The result is the set of  five equations
\begin{subequations}
\label{sixseven}
\begin{eqnarray}
\frac{\partial\langle\pi^i\rangle}{\partial t} & = & -\frac{\partial\langle
T^{ij}\rangle}{\partial x^j}\, ,\label{sixseven a}\\
\frac{\partial\langle\epsilon\rangle}{\partial t} & = & -\vec\nabla\cdot\langle\vec\pi\rangle
\, ,\label{sixseven b}\\
\frac{\partial\langle\nu\rangle}{\partial t} & = & -\vec\nabla\cdot\langle\vec\jmath\rangle\, .
\label{sixseven c}
\end{eqnarray}
\end{subequations}
The expected values are all functions of $\vec x$ and $t$.

The set of equations \eqref{sixseven} close for the following reason:  Eq \eqref{sixfour} with $\tilde\rho_{\rm leq}$ could in principle be inverted to express  $\beta(\vy,t), \vec u(\vy,t), \mu(\vy, t)$, and therefore $\tilde\rho_{\rm leq}$ itself,  in terms of the expected values \eqref{sixone}. Thus the expected values on the right hand side of \eqref{sixseven} become functionals of the quasiclassical variables on the  left hand side and the equations close. 

The process of expression and inversion sketched above could be difficult to carry out in practice.  The familiar classical equations of motion arise from further approximations, in particular from assuming that the gradients of all quantities are small.  For example, for a non-relativistic fluid of particles of mass $m$,
the most general Galilean-invariant form of
the stress tensor that is linear in the gradients of the  fluid velocity  $\vec u(x)$has the approximate form \cite{LL-FM}
\begin{align}
\langle T^{ij} \rangle  =&  p\delta^{ij} + m \nu \, u^i u^j
 - \eta\left[\frac{\partial u^i}{\partial x^j} + \frac{\partial u^j}{\partial x^i} - 
\frac{2}{3}\ \delta_{ij} \left(\vec\nabla \cdot\vec u\right)\right]  \nonumber \\
 &-  \zeta\, \delta_{ij} \left(\vec\nabla \cdot \vec u\right)\, .
\label{sixeleven}
\end{align}
The pressure $p$ and coefficients of viscosity $\eta$ and $\zeta$ are
themselves functions say  of the expected values \eqref{sixone}. This form of the stress tensor in \eqref{sixseven a} leads to the
Navier-Stokes equation.

What determines the volume $V$ defining the coarse-grained variables of the quasiclassical realms?  The volume $V$ must be large enough the ensure the decoherence of histories constructed from these quasiclassical variables, and beyond that to ensure classical predictability in the face of the noise that typical mechanisms of decoherence produce. The volumes must be small enough to allow local equilibrium. Roughly speaking the volume $V$ should be chosen as small as possible consistent with these requirements. That is,  it should be chosen so the  quasiclassical realms are maximally refined consistent with decohence and predictability. Then they are a feature  of our universe and not a matter of our choice. 

We have now removed two more of the deficiencies of the Ehrenfest derivation as shown in the following table:\\

\noindent\shadowbox{\parbox{6.5in}{
\centerline{Deficiencies of the Ehrenfest Derivation}
\begin{itemize}
\item[\chk] {\it Limited to expected values, but classicality is defined through histories.}  

\item[\chk] {\it Deals only with measurements on isolated subsystems with a few degrees of freedom.} 

\item[\chk] {\it Assumes the classical equations follow directly from the fundamental action.} 

\item[\chk] {\it Posits rather than derives the variables which exhibit classical behavior.}  

\item[\cross] {\it Assumes classical spacetime.} 

\end{itemize}
}}

\noindent There remains the origin of classical spacetime to which we turn after a brief discussion of the second law of thermodynamics. 

\section{The Second Law of Thermodynamics}
\label{secondlaw}

The quasiclassical realms of our universe exhibits two important thermodynamic features that are not directly connected to classical determinism:
\begin{itemize}
\item The tendency of a total entropy of the universe to increase.
\item The tendency of this entropy for nearly isolated subsystems to increase in the same direction of time. This may be called the homogeneity of the thermodynamic arrow of time.
\end{itemize}
These two features are connected. The first follows from the second, but only in the late universe when nearly isolated subsystems are actually present. In the early universe we have only the first. Together they may be called the second law of thermodynamics.

Thermodynamics, including the second law, is an essential part of classical physics, and, indeed, a prerequisite for it. In the previous section, for example,  we assumed the second law when we posited the rapid approach to  local equilibrium necessary to derive a closed system of deterministic equations from the conservation relations. 

Entropy is generally a measure of the information missing from a coarse-grained description of a physical system. In the case of the quasiclassical variables \eqref{sixone} we can define it  at a given time as the maximum of the information measure $-Tr(\tilde\rho\log\tilde\rho)$ over density matrices $\tilde\rho$ that preserve the expected values of the quasiclassical variables at that time. More specifically, if $\rho$ is the state of the system, we take
\begin{equation}
S(t) \equiv \underset{\tilde\rho}{\rm max}[-Tr({\tilde\rho}\log{\tilde\rho})],
\label{sl1}
\end{equation}
keeping fixed for each $\vec y$
\begin{equation}
\langle\epsilon_V (\vec y, t)\rangle \equiv Tr (\epsilon_V (\vec y, t) \rho) \approx Tr (\epsilon_V (\vec y, t) \tilde\rho_{\rm eq})\, ,
\label{sl2}
\end{equation}
together with the similar relations for $\vec\pi_V (\vec y, t)$ and $\nu_V (\vec y, t)$. The result is the local equilibrium density matrix \eqref{sixfive}. 

The entropy defined this way {\it is} the usual entropy of chemistry, physics, and statistical mechanics. {\it The coarse-graining in terms of local conserved quantities that exhibits the determinism of the quasiclassical realms thus also defines the entropy for its thermodynamics}.

A special initial quantum state is needed to predict with high probability the classical spacetime whose symmetries are the origin of the conservation laws behind classical determinism. But further conditions on the state are needed for the universe to exhibit the thermodynamic features mentioned above. First, the general increase in total entropy requires that:

\begin{itemize}
\item The quantum state is such that the initial entropy is near the minimum it could have for the coarse graining defining it. It then has essentially nowhere to go but up. 
\item The relaxation time to equilibrium is long compared to the present age of the universe so that the general tendency of its entropy to increase will dominate its evolution.
\end{itemize}  

In our simple model  cosmology we have neglected gravitation for simplicity, but  to understand the origin of the second law it is necessary to consider  it. That is because
gravity is essential to realizing the first of the conditions above. In a self-gravitating system gravitational clumping increases entropy. The matter in the early universe is not clumped and nearly in thermal equilibrium --- already  at maximal entropy. But the spacetime in the early universe is approximately homogeneous, implying that the entropy has much more room to increase through the {\it gravitational} growth of fluctuations. In a loose sense, as far as gravity is concerned, the entropy of the early universe is low for the coarse graining defined by quasiclassical variables.  The entropy then increases. 
The no-boundary quantum state in particular implies that gravitational fluctuations are small in the early universe \cite{HHaw85,HLL93} giving entropy room to grow. 

Coarse graining by approximately conserved quasiclassical variables helps with the second of the two conditions above. 
Small volumes come to local equilibrium quickly. But the approximate conservation ensures that the whole system will approach equilibrium slowly, whether or not such equilibrium is actually attained.  

The homogeneity of the thermodynamic arrow of time, which was the other aspect of the second law mentioned at the beginning of this section, cannot follow from the approximately time-reversible dynamics
and statistics alone. Rather the explanation is that the progenitors of today's nearly isolated
systems were all far from equilibrium  a long time ago and have been running down hill ever since.   As Boltzmann put it
over a century ago: ``The second law of thermodynamics can be proved from the
[time-reversible] mechanical theory, if one assumes that the present state of the
universe$ \dots$ started to evolve from an improbable [{\it i.e.}~special] state''
\cite{Bol97}. There is thus a stronger constraint on the initial state than merely having low total entropy. It must be locally low. 

The initial quantum state of our universe must be such that it leads to the decoherence of
sets of quasiclassical histories  that describe coarse-grained spacetime geometry and matter fields. Our observations require this now, and the successes of the classical history of the universe suggests that there was  a quasiclassical realm at a very early time. In addition,  the initial state must 
be such that the entropy of quasiclassical coarse graining is low in the beginning and also be such  that the entropy of presently isolated systems was also low then. Then the universe can exhibit both aspects of the second law of thermodynamics.

The quasiclassical coarse grainings are therefore distinguished from others, not only
because they exhibit predictable regularities of the universe governed by approximate
deterministic equations of motion, but also because they are characterized by a sufficiently  low entropy in the beginning and a slow evolution towards equilibrium --- two properties which make those regularities exploitable.

\section{The Origin of Classical Spacetime}
\label{classpacetime}

The classical behavior of matter in a given background spacetime depends only weakly on the matter's fundamental quantum physics. The forms of the dyanamical equations
\eqref{sixseven} follow largely from conservation laws and the conditions on the interactions necessary for local equilibrium. In a sense, the quasiclassical realms shield  us from quantum physics --- a happy circumstance that was of great importance historically.

By contrast the origin of classical spacetime is strongly dependent on the physics of quantum gravity and the theory of the initial quantum state of the universe. That is both the attraction of the issue and its difficulty. It is impossible to say much about this in the space made available for this paper. That not least because the quantum theory sketched in Section \ref{dhqm} must be generalized further to deal with quantum spacetime (see, e.g.\cite{Har95c,Har07a}).  The discussion in Sections \ref{dhqm}-\ref{hydro} relied on a fixed notion of time to describe histories --- a notion which is not available when spacetime itself is a quantum variable. The following heuristic discussion may however give some sense of the issues involved. 

Let's first recall one way in which quantum mechanics predicts classical behavior for the motion of a non-relativistic particle. Consider a particle of mass $m$ moving in one dimension $x$ in a potential $V(x)$. Wave functions $\psi(x)$ describe its states. Consider wave  functions that are well approximated in the semiclassical (WKB) form 
\begin{equation}
\psi(x) \approx A(x) \exp[iS(x)/\hbar] 
\label{scpart}
\end{equation}
where $S(x)/\hbar$ varies rapidly with $x$ and $A(x)$ varies slowly. Such states predict classical behavior for the particle.   Specifically they imply that, in a set of alternative histories suitably coarse-grained in $x$ at a series of times, the probabilities are high for correlations in time summarized by the classical equation of motion for the particle (e.g. \cite{HHer09}). 

A wave function satisfying \eqref{scpart} also predicts  probabilities  for {\it which} classical histories satisfy the equation of motion. That is, it predicts probabilities for the initial conditions to the dynamical equations.  Consider histories that pass through a position $x$ at the time the wave function is specified.  Non-zero probabilities are predicted only for the history with momentum $p$ given by
\begin{equation}
p \equiv m \frac{dx}{dt} = -\nabla S(x)  
\label{intcurves}
\end{equation}
and the probability (density)  for this history is $|A(x)|^2$. Thus, a wave function of semiclassical form \eqref{scpart} predicts the probabilities of an  ensemble of classical histories labeled by their initial $x$. 

An analogous discussion of the origin of classical spacetime can be given in quantum cosmology (e.g. \cite{Har95c,HHer09}).
In quantum gravity the metric on spacetime will fluctuate quantum mechanically and generally not behave classically. Consider a simple model in which the quantum metrics are restricted to be homogeneous, isotropic, and spatially closed.  As a model of the matter assume a single homogeneous scalar field $\phi(t)$. 

Spacetime geometry in these models is described by metrics of the form
\begin{equation}
ds^2=-dt^2 +a^2(t) d\Omega^2_3 .
\label{eucmetric}
\end{equation} 
where $d\Omega^2_3$ is the metric on the unit, round, three-sphere. The scale factor $a(t)$ determines how the size of the spatial geometry varies in time. Closed Friedmann-Robertson-Walker cosmological models describing the expansion of the universe from a big bang  have metrics of this form with a scale factor $a(t)$ satisfying the Einstein equation. In quantum mechanics $a(t)$ could have any form. Classical behavior of these minisuperspace models means high probability for coarse-grained $a(t)$'s obeying the Einstein equation.  

A wave function of the universe in this model is a function $\Psi(a,\phi)$ of the scale factor and homogeneous scalar field. Suppose that the wave function in some region of $(a,\phi)$ space is well approximated by the semiclassical form 
\begin{equation}
\Psi(a,\phi) \approx A(a,\phi) \exp[i S(a,\phi)/\hbar] 
\label{scgrav}
\end{equation}
where $S(a,\phi)/\hbar$ is rapidly varying and $A(a,\phi)$ is slowly varying. Then, from the analogy with non-relativistic quantum mechanics, we expect\footnote{See, e.g. \cite{Har95c,HHer09} for a framework in which these expectations are partially borne out.}  the wave function to predict an ensemble of classical spacetimes with initial data related by the analog of \eqref{intcurves} and probabilities related to $|A(a,\phi)|^2$. 

If our universe is a quantum mechanical system, it has a quantum state. A  theory of that state is a necessary part of any `final theory' and the goal of quantum cosmology. Hawking's no-boundary wave function  of the universe \cite{Haw84a} is a leading candidate for this theory. In the context of the simple model the no-boundary wave function is specified by the following functional integral:  
\begin{equation}  
\Psi(a,\phi) =  \int_{\cal C} \delta a' \delta \phi' \exp(-I[a'(\tau),\phi'(\tau)]/\hbar) .
\label{nbwf}
\end{equation}
Here, the path integration is over histories $a'(\tau)$ and $\phi'(\tau)$  of the scale factor and matter field and $I[a'(\tau),\phi'(\tau)]$ is their Euclidean action. The sum is over cosmological geometries that  are regular  on a manifold with only one boundary at which $a'(\tau)$ and $\phi'(\tau)$ take the values $a$ and $\phi$. The integration is carried out along a suitable complex contour ${\cal C}$ which ensures the convergence of \eqref{nbwf} and the reality  of the result.

 Does the no-boundary quantum state predict classical spacetime for the universe and if so what classical spacetimes does it predict?  The answer to the first part of the question   is `yes'. In certain regions of $(a,\phi)$ space the defining  path integral in \eqref{nbwf} can be carried out by the method of steepest descents. The dominent contributions come from the complex extrema of the Euclidean action. The leading order approximation of one extemum is 
\begin{equation}
\Psi(a,\phi) \approx  \exp\{[-I_R(a,\phi) +i S(a,\phi)]/\hbar\}
\label{semiclass}
\end{equation}
where $I_R(a,\phi)$ and $-S(a,\phi)$ are the real and imaginary parts of the Euclidean action evaluated at the extemizing path.

When $S(a,\phi)/\hbar$ varies rapidly and $I_R(a,\phi)\hbar$ varies slowly this is a wave function of the universe of semiclassical form \eqref{scgrav}. An ensemble of classical spacetimes is predicted with different probabilities. The probabilities will be different for such things as whether the universe bounces at a minimum radius or has an initial singularity, how much matter it has, and the duration of an inflationary epoch. These are important issues for cosmology (e.g. \cite{HHH07}). But a quasiclassical realm of matter depends only on the local symmetries of a classical spacetime from the arguments of the preceding three sections.  Each classical spacetime with any matter at all will therefore exhibit  quasiclassical realms. 

Our list of tasks now stands like this:\\

\noindent\shadowbox{\parbox{6.5in}{
\centerline{Deficiencies of the Ehrenfest Derivation}
\begin{itemize}
\item[\chk] {\it Limited to expected values, but classicality is defined through histories.}  

\item[\chk] {\it Deals only with measurements on isolated subsystems with a few degrees of freedom.} 

\item[\chk] {\it Assumes the classical equations follow directly from the fundamental action.} 

\item[\chk] {\it Posits rather than derives the variables which exhibit classical behavior.}  

\item[\chk] {\it Assumes classical spacetime.} 

\end{itemize}
}}

\section{Why we focus on quasiclassical variables}
\label{whydowe}

A quantum universe exhibits many different decoherent sets of alternative coarse-grained histories ---  many different realms. Two realms are {\it compatible} if each one can be fine-grained to yield the same realm. But there are also mutually {\it incompatible realms} for which there is no finer-grained realm of which they are both coarse-grainings \cite{GH90a,GH07}. Quantum mechanics by itself does not prefer any one of these realms over the others. Why then do we as human IGUSes focus almost exclusively on quasiclassical realms?

Questions of the form `Why do we ....?'  can be answered within quantum theory by understanding human IGUSes as particular kinds of physical systems in the universe. As human IGUSes, both individually and collectively, we are described in terms of quasiclassical variables. We are therefore not separate from the universe's quasiclassical realms but rather phenomena exhibited by them\footnote{Are there realms qualitatively different from the quasiclassical ones that exhibit IGUSes? At present we lack a general enough conception of IGUS  to formulate this question precisely much less answer it.}. `Why do we ...?' questions can therefore only be formulated in terms of the probabilities  of the universe's quasiclassical realms and our description within them. 

The elementary answer to the question of why we focus on quasiclassical variables is that we are physical systems described by  quasiclassical variables that possess senses that are adapted to perceive quasiclassical  variables. The predictable regularities of the quasiclassical realms suggests why it is adaptive to have senses that register quasiclassical variables. But collectively we have also evolved to understand and use quantum mechanics. This has also proved adaptive at least in the short run\footnote{Some estimate that a large percentage of the US GDP can be attributed to our understanding of quantum theory. Our understanding of nuclear fission may prove to be less adaptive.}. 

Could the quasiclassical realms of this universe  contain quasiclassically described IGUSes elsewhere whose senses register variables substantially different from the ones we use, even non-quasiclassical ones? 
To answer it would be necessary to calculate the probabilities of alternative evolutionary histories of  such quasiclassically described IGUSes. It is well beyond our power at present to even formulate such a calculation precisely much less carry it out. If we ever encounter extra-terrestrial IGUSes this question may be settled experimentally.

Questions concerning human IGUSes in decoherent histories quantum theory are both fascinating and difficult. But we should emphasize that  answers are not required to understand, utilize, or test the theory for other purposes.  
That is because IGUSes, including human beings, occupy no special  place and play no preferred role this formulation of quantum theory. Rather, they are but one of the many complex systems that can be described within it.

\section{The Copenhagen Approximation}
\label{copenhagen}

Copenhagen quantum mechanics can be seen as an approximation to decoherent histories quantum theory that is appropriate for situations in which a series of measurements is carried out by an apparatus on an otherwise isolated subsystem. 

In Copenhagen quantum mechanics, the isolated subsystem is described quantum mechanically. But the apparatus is described by the separate classical physics posited by the theory. The probabilities for the outcomes of a series of ``ideal'' measurements is given by unitary evolution of the subsystem's state interrupted at the time of measurements by projections onto the values of the outcomes --- the infamous reduction of the wave packet.   

In decoherent histories, apparatus and subsystem are separate parts of one closed system (most generally the universe). In a measurement, 
a variable of the subsystem, perhaps not quasiclassical and perhaps not otherwise decohering, becomes correlated with a quasiclassical variable of an apparatus. 
Histories of the measured variable decohere because of this correlation with the decohering histories of the quasiclassical realm. 

The Copenhagen prescription for the probability of a series of measurement outcomes  can be  derived from the probabilities of decoherent histories quantum theory by modeling the measurement situations to which it applies (e.g.\cite{Har91a}). Idealized  measurement models have a long history in quantum theory (e.g.\cite{LB39}).  A typical model assumes a closed system --- a model universe --- consisting of an apparatus,  a subsystem which it measures, and perhaps other degrees of freedom. The Hilbert space is idealized as a tensor produce ${\cal H}_s \otimes {\cal H}_r$ with the factor ${\cal H}_s$ for the subsystem and the factor ${\cal H}_r$ for the rest including the apparatus. The subsystem is measured by the apparatus at a series of times $t_1, t_2, \cdots, t_n$ and it otherwise isolated  from the rest of the universe. The initial state is assumed to factor into a pure state $|\psi\rangle$ in ${\cal H}_s$ and a density matrix for the rest.

The measurement interaction is idealized to i) occur at definite moments of time, ii) create a perfect correlation between the measured alternatives of the subsystem  and the registrations of the  apparatus --- the former represented by sets of projections $\{s_\alpha(t)\}$ in ${\cal H}_s$ and the latter by projections $\{R_\alpha(t)\}$ in ${\cal H}_r $, and iii) disturb the subsystem as little as possible (an ideal measurement).  Under these assumptions the probability of the sequence of registrations can be shown \cite{Har91a} to be given by 
\begin{equation}
p(\alpha_n, \cdots \alpha_1) = || s^n_{\alpha_n}(t_n) \cdots s^1_{\alpha_n}(t_1) |\psi\rangle||^2 \ . 
\label{cop1}
\end{equation}
The argument of the square in \eqref{cop1} can be thought of as a state of the subsystem which evolved from the initial $|\psi\rangle$  by unitary evolution (constant state in the Heisenberg picture)  interrupted by the action of projections at the times of  measurements (state reduction). This is the usual Copenhagen story.

Eq \eqref{cop1} is a huge and essential simplification when compared to the  basic relation \eqref{dfnal}. Decoherence has been assumed rather than  calculated. More importantly,  \eqref{cop1} refers to a Hilbert space which may involve only a few degrees of freedom whereas \eqref{dfnal} involves all the degrees of freedom in the universe.

Assumptions i)-ii) may hold approximately for many realistic measurement situations. But assumption iii) --- the projection postulate or second law of evolution  --- does not hold for most.\footnote{The idea that the two forms of evolution of the Copenhagen approximation are some kind of problem for quantum theory  seems misplaced from the perspective of the quantum mechanics of closed systems which has no such division.} But it is in this way that Copenhagen quantum mechanics is recovered from the more general decoherent histories quantum mechanics once one has a quasiclassical realm. It is not recovered generally but only for idealized measurement situations. It is not recovered exactly but only to an approximation calculable from the more general theory --- an approximation which is truly excellent for many realistic measurement situations \cite{Har91a}. The separate classical physics posited by Copenhagen quantum theory is an approximation to the quasiclassical realms.  {\it Copenhagen quantum mechanics is thus not an alternative to decoherent histories, but rather contained  within it as an approximation appropriate for idealized measurement situations.}
 
 The founders of quantum mechanics were correct that something besides the wave function and Schr\"odinger equation were needed to understand the theory. But it is not a posited classical world to which quantum mechanics does not apply. Rather, it is the quantum state of the universe together with the theory of quantum  dynamics that explains the origin of quasiclassical realms within the more general quantum mechanics of  closed systems.

\section{Summary and Open Questions}
\label{open}
We now have a complete  sketch of a explanation of the quasiclassical realms in our quantum universe in the context of today's fundamental physics. Our discussion has been top-down --- proceeding from the classical world to the quantum --- starting in today's universe and working backward to the beginning. To summarize we recapitulate these developments from the bottom up.

\begin{itemize}

\item The particular quantum state of our universe implies the classical behavior of spacetime geometry coarse-grained on scales well above the Planck scale. Further,  it predicts the homogeneity of this spacetime  on cosmological scales that implies a low total entropy leading to the second law of thermodynamics.

\item Local Lorentz symmetries of classical spacetime imply conservation of energy and momentum. The effective theory of the matter interactions implies the approximate conservation of various charges and numbers at various stages in the evolution of the universe.

\item Quasiclassical variables specified by ranges of values of the averages of densities of conserved or approximately conserved quantities over small volumes are definable. Sets of alternative histories of these variables decohere and define quasiclassical realms. 

\item When the volumes are suitably large, the approximate conservation of the quasiclassical variables ensure that they evolve predictably despite the noise that typical mechanisms of decoherence produce. 

\item When the volumes are suitably small their contents approach local equilibrium on time scales short compared to those on which the quasiclassical variables are changing. 

\item Local equilibrium implies that the evolution of the quasiclassical variables obeys a closed, deterministic set of equations of motion incorporating constitutive relations determined by  local equilibrium. 

\end{itemize}

The chain above gives a broad outline of how the quasiclassical realms of our universe emerge from its fundamental quantum physics and particular quantum state. However, touch this chain where you will and there are issues that remain to make it more realistic, more general, more complete, more precise, and more quantitative. The following is a short and selective list of outstanding problems:

{\it Decoherence of  Classical Spacetime:}  Our understanding of the {\it emergence of} classical spacetime from  particular states in quantum gravity is more primitive than our understanding of the emergence of the classical behavior of matter {\it given} a fixed spacetime. Even a cursory comparison of Section \ref{classpacetime} with Section \ref{hydro} reveals this. Partly this is because we lack a complete and manageable quantum theory of gravity. But even in the low energy effective theory of gravity based on general relativity we do not have precise notions of the diffeomorphism invariant coarse grainings\footnote{Defining diffeomorphism invariant coarse grainings of matter fields in quantum spacetime is itself an issue, see. e.g. \cite{GMH06}.} that {\it define} the classical behavior of geometry in every day situations above the Planck scale. And, perforce, we have an inadequate understanding of the mechanisms effecting their decoherence. 

{\it More Realistic Models:} The model universe of a static box of particles interacting by short range potentials that was discussed all too briefly in Section \ref{hydro} is highly simplified. Models are needed which incorporate at least the following features of the realistic universe.

\begin{itemize}
\item{\it Cosmology:}  The expansion of the universe, gravitational clumping, possible eternal inflation, the decay of the proton, the  formation and evaporation of black holes. 

\item{\it Degrees of Freedom:}  The relativistic quantum fields that are the basic variables of today's effective field theories.

\item{\it Coarse-graining:} Branch dependent\footnotemark[9] coarse-grainings that express narratives directly in terms of realistic hydrodynamic variables. 

\item{\it Maximal Refinement:}  Maximal refinement of  coarse-grainngs  consistent with decoherence and classicality  so that the quasiclassical realms are a feature of the universe and not a matter of human choice as discussed at the end of Section \ref{hydro}.

\item{\it Initial States:} Initial states that arise from theories of the quantum state of the universe and not from ad hoc assumptions about an environment as in \eqref{denmatrix}. 

\end{itemize}

{\it Comparing Different Realms:} As mentioned in Section \ref{whydowe}, a quantum universe can be described by many decohering sets of alternative coarse-grained histories --- many realms. The quasiclassical realms are distinguished by a high level of classical predictability and a low initial entropy among other properties. Intuitively they provide the simplest description of the general regularities of the universe that are readily exploitable by IGUSes of the kinds we know about.
A genuine comparison of the quasiclassical realms with others the universe exhibits would require quantitative measures on realms of simplicity, predictability, classicality, etc. Various approaches to such measures have been explored  \cite{GH94}  but  no complete satisfactory result has yet emerged. 

Thus while we have gone far beyond the Ehrenfest derivation, there is still a long way to go! 

\section{Quasiclassical Realms and Fundamental Physics}
\label{conclusions}
From the present theoretical perspective, a final theory consists of two parts: (1) a dynamical theory specifying quantum evolution (the Hamiltonian in simple models), and (2) a theory of the universe's quantum state. Without both there are no predictions of any kind. With both,  probabilities for the members of  every decoherent  set of alternative histories of the univerese are in principle predicted. 

Today the search for a final theory has taken physics further and further from the determinism and unique reality that characterized classical physics. A final theory may incorporate quantum indeterminacy, mutually incompatible realms, and not have spacetime at a basic level. In that context, the seemingly prosaic quasiclassical realms of our universe appear remarkable. 

On what features of the two parts of a final theory do the quasiclassical realms depend? The discussion in this essay suggests the following:
 
{\it Requirements for Dynamics:} For the most part, what is required of the dynamical part of theory is an effective theory of the elementary interactions  which has the properties necessary for local equilibrium at the matter energies well below the Planck scale that are reached in an expanding universe. Specifically, the interactions should be approximately local and dominantly short range.

However, the specific properties of the  only  unscreened long range interaction --- gravity --- are crucial for the quasiclassical realms. It is the gravitationally driven expansion of the universe that ensures the separation of the energy scales of matter from those of quantum gravity. It is the attractive and universal character of gravity which allows isolated systems to form by the growth and collapse of fluctuations. And it is the relative weakness of the gravitational interaction which allows the universe to remain out of total equilibrium on the time scale of its present age. 
 
{\it Requirements for the State:}  More is required of the initial state. It must be such as to imply that histories of cosmological geometry coarse-grained above the Planck scale behave classically. The local symmetries of this classical spacetime imply conservation laws which determine in part the variables characterizing the quasiclassical realms.  
Further, the quantum state must imply an initial condition of low total entropy so that the universe can exhibit the second law of thermodynamics

Almost as important as what the quantum state is required to predict is what it is not required to predict. The beauty of quantum theory is that probabilities are basic. A simple  discoverable theory of the quantum state is therefore unlikely to predict with high probability the {\it particular} classical history we observe with all its apparent complexity. Rather it predicts the simple dynamical regularities common to every classical  history with high probability,  leaving to quantum accidents the complexity of particular configuration of matter observed. Thus quantum mechanics allows the laws determining probabilities to be simple and still be consistent with present complexity.  

It is possible to emphasize how specific these requirements for a quasiclassical realm are. Surely they will not be satisfied by every  state in quantum gravity nor every conceivable theory of quantum dynamics. They are sufficiently specific that classicality could be important as a vacuum state selection principle \cite{HHH07} in theories like string theory that permit many.

However, it is equally striking how little is required for a final theory to exhibit a quasiclassical realm. The small number and general nature of the requirements discussed above mean that there must be many states and dynamical theories that manifest a quasiclassical realm. Indeed, historically classical physics has shielded us from the nature of the final theory. Given classical spacetime,  the form of the classical equations of motion was determined by conservation laws plus Maxwell's equations for the electromagnetic field and the Einstein equation for spacetime geometry. The equations of state, susceptabilities, etc that entered into these equations could be determined phenomenologically. It was thus not necessary even to know about atoms much less their quantum mechanics to explore classical regularities. As far as quantum gravity is concerned, the expansion of the universe  has shielded us from an immediate need to consider it by driving the characteristic scales of  matter away from the Planck scale.

In these ways our particular universe has allowed a step by step, level by deeper level journey of discovery of the fundamental regularities --- a journey which we have not yet completed. The quasiclassical realms of every day experience have played a central role in this journey, both as a starting point for the exploration and as the chief observational feature of our quantum universe to be explained.  

\acknowledgments

Discussions with  Murray Gell-Mann and Jonathan Halliwell  over many years  on the emergence of classical behavior in the universe are gratefully acknowledged.  This work was supported in part by the National Science Foundation under grant PHY05-55669.


\renewcommand{\theequation}{\Alph{section}.\arabic{equation}}

\begin{appendix}
\section{The Quantum Mechanics of Closed Systems}

Largely to explain the notation this appendix  gives a bare-bones account of some essential elements of the modern synthesis of ideas constituting  the decoherent histories quantum mechanics of closed systems \cite{Gri02, Omn94, Gel94}. 
 
The most general objective of quantum theory is the prediction of the probabilities of
individual members of sets of coarse-grained alternative histories of the closed
system. For instance, we might be interested in alternative histories of the center-of-mass
of the Earth in its progress around the Sun, or in histories of the correlation between the
registrations of  a measuring apparatus and a  property of the subsystem. Alternatives at one moment of
time can always be reduced to a set of yes/no questions.  For example, alternative positions 
of the Earth's center-of-mass can be reduced to asking, ``Is it in this region -- yes or no?'',
``Is it in that region -- yes or no?'', etc. An exhaustive set of yes/no alternatives at one time is
represented in the Heisenberg picture by an exhaustive set of orthogonal projection operators 
$\{P_\alpha(t)\}$,
$\alpha = 1, 2, 3 \cdots$.  These satisfy
\begin{equation}
\sum\nolimits_\alpha P_\alpha(t) = I, \  {\rm and}\ P_\alpha(t)\, P_\beta (t) =
\delta_{\alpha\beta} P_\alpha (t) \ ,
\label{twoone}
\end{equation}
showing that they represent an exhaustive set of exclusive alternatives.  In the Heisenberg
picture, the operators $P_\alpha(t)$ evolve with time according to 
\begin{equation}
P_\alpha(t) = e^{+iHt/\hbar} P_\alpha(0)\, e^{-iHt/\hbar}\, .
\label{twotwo}
\end{equation}
The state $|\Psi\rangle$ is unchanging in time.

An important kind of set of histories is specified by a series\footnote{Realistically the sets are {\it branch dependent} with the sets at one time depending on the particular sequence of alternatives at preceding times. However we ignore branch dependence in this simplified exposition.} of sets of single time
alternatives $\{P^1_{\alpha_1} (t_1)\}$,
$\{P^2_{\alpha_2} (t_2)\}, \cdots$, $\{P^n_{\alpha_n} (t_n)\}$ at a sequence of times
$t_1<t_2<\cdots < t_n$.  The sets at distinct times can differ and are distinguished by the
superscript on the $P$'s. For instance, projections on ranges of position might be followed
by projections on ranges of momentum, etc.  An individual history $\alpha$ in such a set is a
particular sequence of alternatives $(\alpha_1, \alpha_2, \cdots, \alpha_n)\equiv \alpha$
and is represented by the corresponding chain of projections called a {\it chain or
class operator}
\begin{equation}
C_\alpha\equiv P^n_{\alpha_n} (t_n) \cdots P^1_{\alpha_1} (t_1)\, .
\label{twothree}
\end{equation}

A set of histories like one specified by \eqref{twothree} is generally {\it coarse-grained}  because alternatives are specified
at some times and not at every time and because the alternatives at a given time are typically 
projections on subspaces with dimension greater than one and not projections onto a complete
set of states. Perfectly  {\it fine-grained} sets of histories consist of one-dimensional
projections at each and every time.

Operations of fine and coarse graining may be defined  on sets of histories.  A set of histories $\{\alpha\}$ may
be {\it fine -grained} by dividing up each class into
 an exhaustive set of exclusive subclasses $\{ \alpha' \}$.  Each subclass  consists of some number of histories in a coarser-grained class, and
every finer-grained subclass is in some  class.   {\it Coarse graining} is the operation of uniting subclasses of histories into bigger classes.  Suppose, for example, that the position
of the Earth's center-of-mass is specified by dividing space into cubical regions of a
certain size. A coarser-grained description of position could consist of
larger regions made up of unions of the smaller ones. Consider a set of histories with class operators $\{C_\alpha \}$ and a coarse graining with class operators
 $\{ {\bar C}_{\bar\alpha} \}$  . The operators $\{{\bar C}_{\bar\alpha}\}$  are then related to the  operators $\{C_\alpha\}$ by summation, {\it viz.}
\begin{equation}
\bar C_{\bar\alpha} = \sum_{\alpha\in\bar\alpha} C_\alpha \ ,
\label{twofour}
\end{equation}
where  the sum is over the $C_\alpha$ for all finer-grained histories $\alpha$  contained within $\bar\alpha$.

For any individual history $\alpha$, there is a {\it branch state vector}  defined by
\begin{equation}
|\Psi_\alpha\rangle = C_\alpha |\Psi\rangle\, .
\label{twofive}
\end{equation}
When probabilities can be consistently assigned to the individual histories in a set,
they are given by
\begin{equation}
p(\alpha) = \parallel |\Psi_\alpha\rangle\parallel^2 = 
\parallel C_\alpha |\Psi\rangle\parallel^2 = ||P^n_{\alpha_n} (t_n) \cdots P^1_{\alpha_1} (t_1)|\Psi\rangle||^2 .
\label{twosix}
\end{equation}
However, because of quantum interference,  probabilities cannot be consistently assigned to every set of alternative
histories that may be described.  The two-slit
experiment  provides an elementary example: An electron emitted by a source can pass
through either of two slits on its way to detection at a farther screen.  It would be
inconsistent to assign probabilities to the two histories distinguished by which slit the
electron goes through if no ``measurement'' process determines this.  Because of interference, the probability for arrival  at a point
on the screen would not be the sum of the probabilities to arrive there by going
through each of the slits. In quantum theory, probabilities are squares of amplitudes
and the square of a sum is not generally the sum of the squares.

Negligible interference between the branches of a set
\begin{equation}
\langle\Psi_\alpha|\Psi_\beta\rangle\approx 0 \quad , \quad \alpha\not=\beta \, ,
\label{twoseven}
\end{equation}
is a sufficient condition for the probabilities \eqref{twosix} to be consistent with the
rules of probability theory. The orthogonality of the branches is approximate in
realistic situations. But we mean by \eqref{twoseven} equality to an accuracy that defines probabilities well beyond the
standard to which they can be checked or, indeed, the physical situation modeled
\cite{Har91a}. 

Specifically, as a consequence of \eqref{twoseven}, the probabilities \eqref{twosix} obey
the most general form of the probability sum rules
\begin{equation}
p(\bar\alpha) \approx \sum_{\alpha\in\bar\alpha} p(\alpha)
\label{twoeight}
\end{equation}
for any coarse graining $\{{\bar\alpha}\}$ of the $\{\alpha\}$.  Sets of
histories obeying \eqref{twoseven} are said to (medium) decohere. 
Medium-decoherent sets are thus the ones for which quantum mechanics makes predictions of consistent  probabilities through 
\eqref{twosix}.  The decoherent sets exhibited by our universe are determined through \eqref{twoseven} and by the Hamiltonian $H$ and the quantum state $|\Psi\rangle$.
The term {\it realm} is used as a synonym for a decoherent set of coarse-grained
alternative histories.

An important mechanism of decoherence is the dissipation of phase coherence between
branches into variables not followed by the coarse graining.  Consider by way of
example, a dust grain in a superposition of two positions deep in interstellar space
\cite{JZ85}.  In our universe, about $10^{11}$ cosmic background photons scatter from
the dust grain each second.  The two positions of the grain become correlated with different, nearly
orthogonal states of the photons. Coarse grainings that follow only the position of the
dust grain at a few times therefore correspond to branch state vectors that are nearly
orthogonal and satisfy \eqref{twoeight}.  

Measurements and observers play no fundamental role in this general formulation
of usual quantum
theory.  The probabilities of measured outcomes can be computed and are given to
an excellent approximation by the usual story. But, in a set of histories where they
decohere, probabilities can be assigned to the position of the Moon when it is not
receiving the attention of observers and to the values of density fluctuations in the early 
universe when there were neither measurements taking place nor observers to carry them out.
\end{appendix}

\end{document}